\documentclass[twocolumn,prb,superscriptaddress]{revtex4-2}

\usepackage[T1]{fontenc}
\usepackage[utf8]{inputenc}
\usepackage{lmodern}
\usepackage{graphicx}
\usepackage{amsmath}
\usepackage{xcolor}
\usepackage{float}
\usepackage{hyperref}
\usepackage{braket}
\usepackage{amssymb,amsmath,amsthm, array}
\usepackage{mdframed}
\usepackage{systeme,mathtools}
\usepackage{yfonts}
\usepackage{comment}
\usepackage{graphicx,psfrag,color,dsfont}
\usepackage{slashed,amssymb,mathtools}

\DeclareUnicodeCharacter{0301}{-}

\def\u{\uparrow}
\def\nn{\nonumber}

\begin{document}

\title{Entanglement between quantum dots transmitted via Majorana wire: \\
     Insights from the fermionic negativity, concurrence and quantum mutual information}

\author{C. Jasiukiewicz}
\affiliation{Department of Physics and Medical Engineering, Rzesz\'ow University of Technology, 35-959 Rzesz\'ow, Poland}
\author{A. Sinner}
\affiliation{Institute of Physics, University of Opole, 45-052 Opole, Poland}

\author{I. Weymann}
\affiliation{Institute of Spintronics and Quantum Information,
Faculty of Physics and Astronomy,
A. Mickiewicz University, 61-614 Poznań, Poland}
\author{T. Doma\'nski}
\affiliation{Institute of Physics, M. Curie-Sk\l{}odowska University, 20-031 Lublin, Poland}
\author{L. Chotorlishvili}
\affiliation{Department of Physics and Medical Engineering, Rzesz\'ow University of Technology, 35-959 Rzesz\'ow, Poland}


\date{\today}
\begin{abstract}
We study quantum entanglement in a system comprising two quantum dots interconnected through the short topological superconducting nanowire, which hosts overlapping boundary Majorana modes. Inspecting the fermionic negativity, we analyze the variation of entanglement
against the position of the energy levels of quantum dots and their hybridization
with the topological superconducting nanowire. In the absence of electron correlations,
the optimal entanglement occurs when the energy levels coincide with the zero-energy Majorana modes, whereas upon increasing the hybridizations, the entanglement is gradually suppressed.
Such monotonous behavior is no longer valid when the quantum dot levels are detuned from
the zero-energy. Under these circumstances, the quantum dots become maximally entangled for a certain optimal hybridization.
Moreover, we study the thermal \textit{concurrence} to explore the entanglement properties at finite temperatures. 
We also compute the \textit{quantum mutual information} and propose recipes for robust finite-temperature entanglement transmission via Majorana modes. 
\end{abstract}
\maketitle

\section{Introduction}

Quantum entanglement is an imaginary physical quantity that characterizes the degree of quantumness of a physical system, and it has been widely studied over the last three decades \cite{RevModPhys.81.865,PhysRevLett.80.2245,PhysRevLett.78.2275,PhysRevLett.98.110502,bruss2002characterizing,RevModPhys.80.517}. 
For practical reasons, the entanglement is interesting because of several unique properties and facets that it exhibits. 
Its analysis provides us with a deeper understanding of the nature of quantum states \cite{PhysRevLett.130.100202,PhysRevB.108.134411,PhysRevB.106.224418},
and besides that, the entangled states can be viewed as a repository of quantum information. For instance, in resource theory, the coherence of entangled states is regarded as means for performing operations on gates, that would not be available in the classical case \cite{RevModPhys.89.041003}. 
Since very recently, research has focused on the entanglement between fermionic degrees of freedom \cite{PhysRevB.97.165123,Dagotto_2024}.
In particular, it was discovered that the fermionic entanglement may exceed the bosonic one.
In addition, the fermionic entanglement is stronger.
In contrast to the bosonic case, the fermionic negativity decays as a power law \cite{PhysRevResearch.6.023125,PhysRevD.109.L071501,PhysRevA.80.012325}. 

\begin{figure}[t]
	\includegraphics[width=0.99\columnwidth]{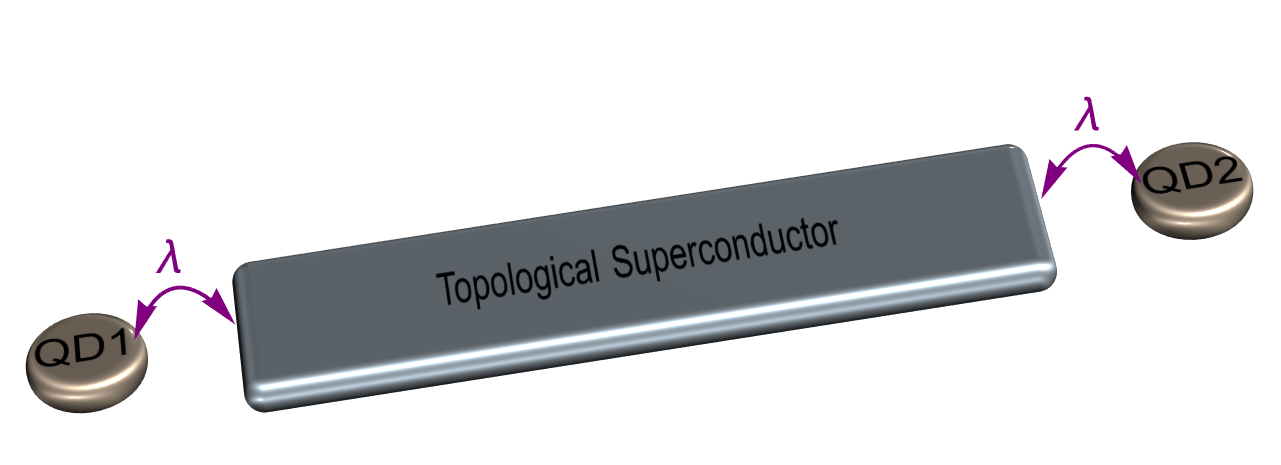}
	\caption{
		The schematic of the investigated setup with symmetric
		coupling $\lambda$ between the quantum dots (QDs)
		and topological superconducting nanowire, 
		with Majorana modes located at its ends.
	}
	\label{fig:device}
\end{figure}

The problem of the quantum entanglement quantification has engaged
the physicists ever since its introduction in the early days of quantum mechanics \cite{EPRS1935,Schroedinger1935}.
In this article we in particular investigate the key entities of the quantum information science,
including the fermionic logarithmic entanglement negativity (or simply the logarithmic negativity), quantum concurrences, and mutual information, in order to shed light on the properties of superimposed quantum states dwelling in a topological superconducting nanowire.
We especially focus on the regime where the Majorana modes,
i.e. the zero-energy quasiparticles appearing at the edges of topologically nontrivial superconducting systems, emerge \cite{RevModPhys.87.137,beenakker2013search,kitaev2001unpaired,alicea2012new, PhysRevB.84.144522}.
Such quasiparticles do always exist in pairs, but it is not obvious to what extent 
are they mutually cross-correlated. This issue is particularly relevant for recent realizations of a minimal Kitaev chain, using two \cite{dvir2023realization} and three \cite{PhysRevLett.132.056602} quantum dots contacted by superconductor where Majorana modes might partly overlap with one another. To inspect their entanglement, we study a hybrid system,
consisting of two quantum dots (QDs) interconnected through Majorana quasiparticles \cite{PhysRevB.109.075432, PhysRevB.84.201308, PhysRevB.84.140501}. A particular configuration of such a system is shown schematically in Fig.~\ref{fig:device}.
In what follows, we investigate the long-range entanglement between these quantum dots transmitted solely by the Majorana modes \cite{Souto_2023}. 
In practical terms, our setup represents a tripartite system
with the corresponding collective density matrix $\hat\rho_{d_1d_2f}$,
where the indices $d_1,\,d_2$ refer to both quantum dots, respectively,
which communicate with each other indirectly through the Majorana fermions
that are described by the third part $f$.
We consider the partial transpose of the reduced density matrix of quantum dots and use the measure of logarithmic negativity to quantify the entanglement in the system. 
Each fermionic species lives on a two-dimensional Fock space, which reflects the only empty and filled state allowed by the fermionic statistics.
Hence, we essentially deal with a three-qubit system, on general properties of which only a patchwork knowledge exists. For instance, 
the separability properties of a generic three-qubit density matrix have been investigated rigorously in Ref.~\cite{Werner2001}. 
Crucially, because of the fermionic anticommuting algebra, the partial transpose
for the fermionic case is less trivial as compared to the bosonic one \cite{eisler2015partial}.
Here, we follow the partial time-reversal transformation method developed in Ref.~\cite{PhysRevB.95.165101},
which allows for capturing all essential topological features of Majorana fermions from studies of the logarithmic negativity. 
A considerable progress has been achieved recently with regard to the non-local entanglement properties of the Majorana-bound states. 
For instance, in Ref.~\cite{PhysRevB.110.224510} the authors studied for the first time two specific entanglement measures for zero on-site energies of the QDs, namely the concurrence and quantum discord. In the present work, we explore fermionic negativity for arbitrary energies of QDs and analyze entanglement at finite temperatures via the thermal concurrence and quantum mutual information.

\section{Model and effective Hamiltonian}

The low-energy physics of the system comprising two quantum dots
connected by a short nanowire composed of a topological superconducting material 
(cf. Fig.~\ref{fig:device}) is captured by the following effective Hamiltonian
\begin{eqnarray}\label{Total Hamiltonianone}
\hat H=\sum\limits_{i=1,2}\hat H^{QD}_{i}+\hat V.
\end{eqnarray}
The QDs are treated as single level spin-less impurities 
\begin{eqnarray}
\label{Total Hamiltonianonetwo}
\hat H^{QD}_{i}=\varepsilon_{i}\hat d^\dag_{i}\hat  d_{i},
\end{eqnarray}
where $\hat  d^\dag_{i}$, $\hat  d_{i}$ are the creation and annihilation operators
of the electrons of energy $\varepsilon_{i}$ in the $i$-th quantum dot.
The quantum dots are hybridized with the Majorana quasiparticles via
\begin{eqnarray}
\hat V = i\varepsilon_M\hat\gamma_1\hat\gamma_2 
+ \lambda_1\!\left(\hat d^\dag_{1}-\hat  d_{1}\right)\! \hat\gamma_1+
i\lambda_2\hat\gamma_2\! \left(\hat  d^\dag_{2}+\hat  d_{2}\right) .
\label{Total Hamiltonianonethree}
\end{eqnarray}
We assume that the electrons of the quantum dots are coupled
to these Majorana boundary modes with the coupling strength $\lambda_i$, respectively.
The overlap $\varepsilon_M$ between the Majorana modes
is finite when the topological superconducting nanowire is shorter
than the superconducting coherence length.
On the other hand, the self-hermitian operators $\hat\gamma_i=\hat\gamma^\dag_i$
can be expressed in terms of conventional fermionic operators $\hat f^\dag$ and $\hat f$,
which obey the standard anticommutation rules, as their superpositions 
\begin{eqnarray}
\hat\gamma_1 = \frac{1}{\sqrt{2}}(\hat f^\dag+\hat f), &\;\;\;&
\hat\gamma_2 = \frac{i}{\sqrt{2}}(\hat f^\dag-\hat f).
\end{eqnarray}

We note that the above Hamiltonian describes the system
at low energies, assuming large Coulomb correlations and magnetic field,
which is needed to induce the topological superconductivity.
This effectively implies that only one spin species,
which couples to the Majorana quasiparticles,
is relevant in the quantum dots
\cite{PhysRevB.84.140501,PhysRevB.84.201308}.

To start our discussion of \textit{negativity},
we consider the following occupation number basis
$|n^{}_f,n^{}_{d^{}_1},n^{}_{d^{}_2}\rangle$,
enumerating the basis vectors as: 
\begin{eqnarray}
\nn\displaystyle
\ket{\varphi^{}_1}=\ket{1,0,0}, &\hspace{0.3cm}&
\displaystyle\ket{\varphi^{}_2}=\ket{0,0,1}, \\
\nn\displaystyle
\ket{\varphi^{}_3}=\ket{0,1,0}, &\hspace{0.3cm}& 
\displaystyle\ket{\varphi^{}_4}=\ket{1,1,1}, \\
\nn\displaystyle
\ket{\varphi^{}_5}=\ket{0,0,0}, &\hspace{0.3cm}&
\displaystyle\ket{\varphi^{}_6}=\ket{1,0,1}, \\
\nn\displaystyle
\ket{\varphi^{}_7}=\ket{1,1,0}, &\hspace{0.3cm}&
\displaystyle\ket{\varphi^{}_8}=\ket{0,1,1}. 
\end{eqnarray}
Then using the algebra of fermionic operators
\begin{subequations}
\begin{eqnarray}
\label{eq:Fixing}
|1,1,1\rangle &=& \hat f^\dag \hat d_1^\dag \hat d_2^\dag |0,0,0\rangle, \\
|0,0,0\rangle &=&  \hat d_2 \hat d_1 \hat f |1,1,1\rangle =   \hat d_2 \hat d_1 \hat f \hat f^\dag \hat d_1^\dag \hat d_2^\dag |0,0,0\rangle, \hspace{5mm}
\end{eqnarray}
\end{subequations}
as well as the explicit form of the interaction term 
\begin{eqnarray}
\nonumber
\hat V &=& \frac{\varepsilon_M}{2}(\hat f^\dag \hat f -\hat  f \hat f^\dag) \\
\nonumber
&-& \frac{\lambda^{}_1}{\sqrt{2}}(\hat f^\dag \hat d^\dag_{1} + \hat f \hat d^\dag_{1} + \hat d^{}_{1}\hat  f^\dag + \hat d^{}_{1} \hat f) \\
&-& \frac{\lambda^{}_2}{\sqrt{2}}(\hat f^\dag \hat d^\dag_{2} - \hat f \hat d^\dag_{2} - \hat d^{}_{2} \hat f^\dag + \hat d^{}_{2} \hat f),
\end{eqnarray}
we construct the Hamiltonian matrix 
\begin{eqnarray}\label{Matrix}
&&\begin{bmatrix}
\frac{\varepsilon{}_M}{2} & -\frac{\lambda^{}_2}{\sqrt{2}}  & \frac{\lambda^{}_1}{\sqrt{2}} & 0 & 0 &  0 & 0 & 0 \\
-\frac{\lambda^{}_2}{\sqrt{2}} & H_{55}  & 0 & -\frac{\lambda^{}_1}{\sqrt{2}}  & 0 &  0 & 0 & 0 \\
\frac{\lambda^{}_1}{\sqrt{2}}  & 0 & H_{33} &  \frac{\lambda^{}_2}{\sqrt{2}} & 0 & 0 & 0 & 0 \\
 0 & -\frac{\lambda^{}_1}{\sqrt{2}} & \frac{\lambda^{}_2}{\sqrt{2}} & H_{88} & 0 & 0 & 0 & 0 \\
0 & 0 & 0 & 0 &  -\frac{\varepsilon{}_M}{2} & -\frac{\lambda^{}_2}{\sqrt{2}} & -\frac{\lambda^{}_1}{\sqrt{2}} & 0 \\
0 & 0 & 0 & 0 & -\frac{\lambda^{}_2}{\sqrt{2}}  & H_{66}  & 0 & \frac{\lambda^{}_1}{\sqrt{2}}\\
0 & 0 & 0 & 0 & -\frac{\lambda^{}_1}{\sqrt{2}} &  0 & H_{44} & \frac{\lambda^{}_2}{\sqrt{2}}\\
0 & 0 & 0 &  0 & 0 & \frac{\lambda^{}_1}{\sqrt{2}} &  \frac{\lambda^{}_2}{\sqrt{2}}  & H_{77} 
\end{bmatrix},
\end{eqnarray} 
which has a block-diagonal structure due to different parities.
Here, the basis vectors are arranged according to the charge parity as follows
\begin{equation}
\underbrace{\ket{\varphi^{}_1},\ket{\varphi^{}_2},\ket{\varphi^{}_3},\ket{\varphi^{}_4}}_{\rm odd\ parity}\;,\;
\underbrace{\ket{\varphi^{}_5},\ket{\varphi^{}_6},\ket{\varphi^{}_7},\ket{\varphi^{}_8}}_{\rm even\ parity}.
\end{equation}
In Eq.~(\ref{Matrix}) we use the following shorthand notation:
\begin{eqnarray}
\nn&\displaystyle
H_{33}=\varepsilon^{}_{1}-\frac{\varepsilon_M}{2},\; 
H_{44}=\varepsilon_{1}+\frac{\varepsilon_M}{2}, \;&\\
\nn&\displaystyle
H_{55}=\varepsilon_{2}-\frac{\varepsilon_M}{2},\;
H_{66}=\varepsilon_{2}+\frac{\varepsilon_M}{2},\;&\\
\nn&\displaystyle
H_{77}=\varepsilon_{1}+\varepsilon_{2}-\frac{\varepsilon_M}{2},\; 
H_{88}=\varepsilon_{1}+\varepsilon_{2}+\frac{\varepsilon_M}{2}.&
\end{eqnarray}
We also note that treating the Coulomb repulsion, appearing in form of an additional Hubbard interaction term, in a mean-field fashion
would merely rescale the QDs energies in the above formulas. We address this issue in more details in Appendix~\ref{app:Coulomb}.

\section{Reduced matrix elements}

We diagonalize the Hamiltonian (\ref{Matrix}) numerically and find the eigenstates 
\begin{equation}
\ket{G_m}=\sum\limits_{n=1}^8C_{mn}\ket{\varphi^{}_n}. 
\end{equation}
The decomposition coefficients $C_{mn}[\varepsilon_M,\varepsilon_{1},\varepsilon_2, \lambda_{1},\lambda_2]$
are functions of several parameters. At zero temperature,
the density matrix of the total system is a pure state and has the form $\hat\rho=\ket{G}\bra{G}$,
where $\ket{G}$ is one from the eigenstates (hereafter we omit the second index $m$).
After tracing the Majorana states $\ket{n_f}$, we obtain the reduced density matrix $\hat\rho^{}_R$
\begin{eqnarray}\label{Matrixreduced}
\hat\rho^{}_R=\begin{bmatrix}
\rho_{11} & \rho_{12} & \rho_{13} & \rho_{14} \\
\rho_{21} & \rho_{22} & \rho_{23} & \rho_{24} \\
\rho_{31} & \rho_{32} & \rho_{33} & \rho_{34} \\
\rho_{41} & \rho_{42} & \rho_{43} & \rho_{44}
\end{bmatrix}.
\end{eqnarray}
The elements of the reduced matrix are given by the following equations: 
\begin{eqnarray}
\nn
\displaystyle \rho_{12}=\rho_{21}^*=C_1C_7^*+C_5C_3^*, &&\;\;
\displaystyle \rho_{13}=\rho_{31}^*=C_1C_6^*+C_5C_2^*, \\
\nn
\displaystyle\rho_{14}=\rho_{41}^*=C_1C_4^*+C_5C_8^*, &&\;\;
\displaystyle\rho_{23}=\rho_{32}^*=C_3C_2^*+C_7C_6^*, \\
\nn
\displaystyle\rho_{24}=\rho_{42}^*=C_3C_8^*+C_7C_4^*, &&\;\;
\displaystyle\rho_{34}=\rho_{43}^*=C_2C_8^*+C_6C_4^*, \\
\nn
\displaystyle\rho_{11}=|C_1|^2+|C_5|^2, &&\;\; \displaystyle\rho_{22}=|C_3|^2+|C_7|^2, \\
\nn
\displaystyle\rho_{33}=|C_2|^2+|C_6|^2, &&\;\;\displaystyle\rho_{44}=|C_4|^2+|C_8|^2 ,
\end{eqnarray}
with the sum of diagonal elements fixed to the unity. The corresponding basis of vectors $\ket{n_{d_1},n_{d_2}}$ is defined as
\begin{eqnarray}
\nn
\displaystyle\ket{\psi^{}_1}=\ket{0}_1\ket{0}_2, &&\;\; \displaystyle\ket{\psi^{}_2}=\ket{1}_1\ket{0}_2, \\
\nn
\displaystyle\ket{\psi^{}_3}=\ket{0}_1\ket{1}_2, &&\;\; \displaystyle\ket{\psi^{}_4}=\ket{1}_1\ket{1}_2.
\end{eqnarray}

\begin{figure*}[t]
\includegraphics[width=0.32\textwidth]{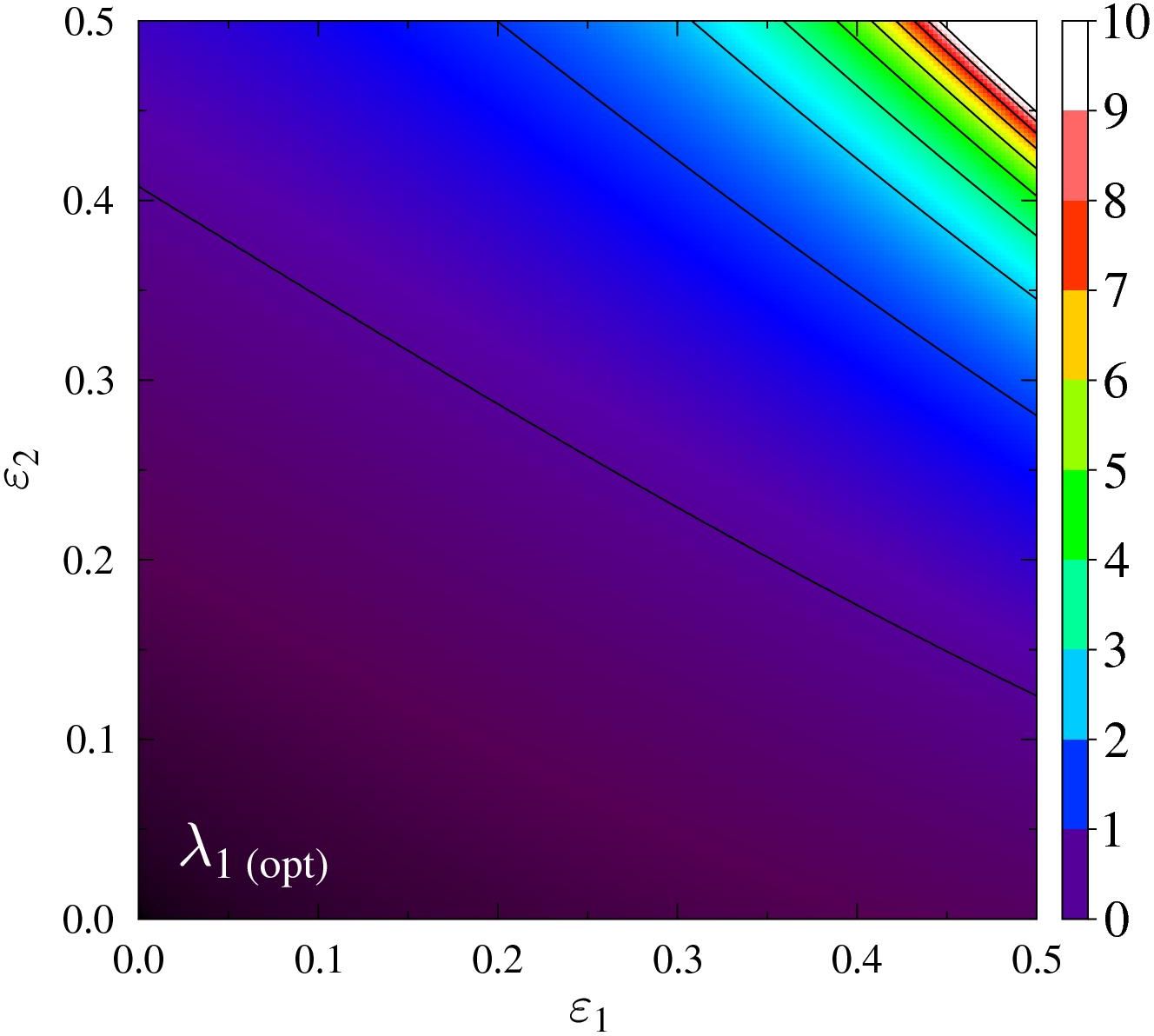}
\includegraphics[width=0.32\textwidth]{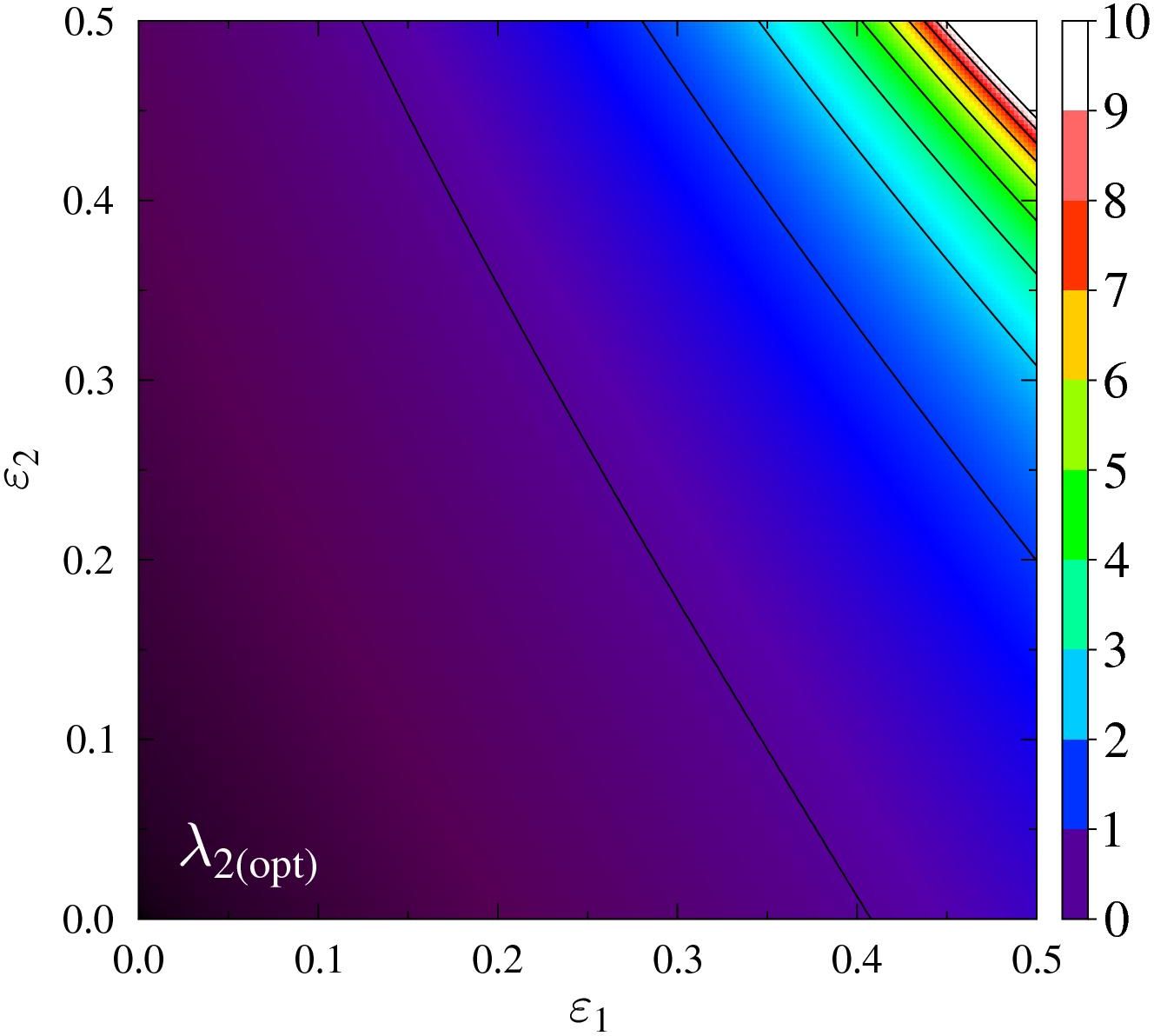}
\includegraphics[width=0.33\textwidth]{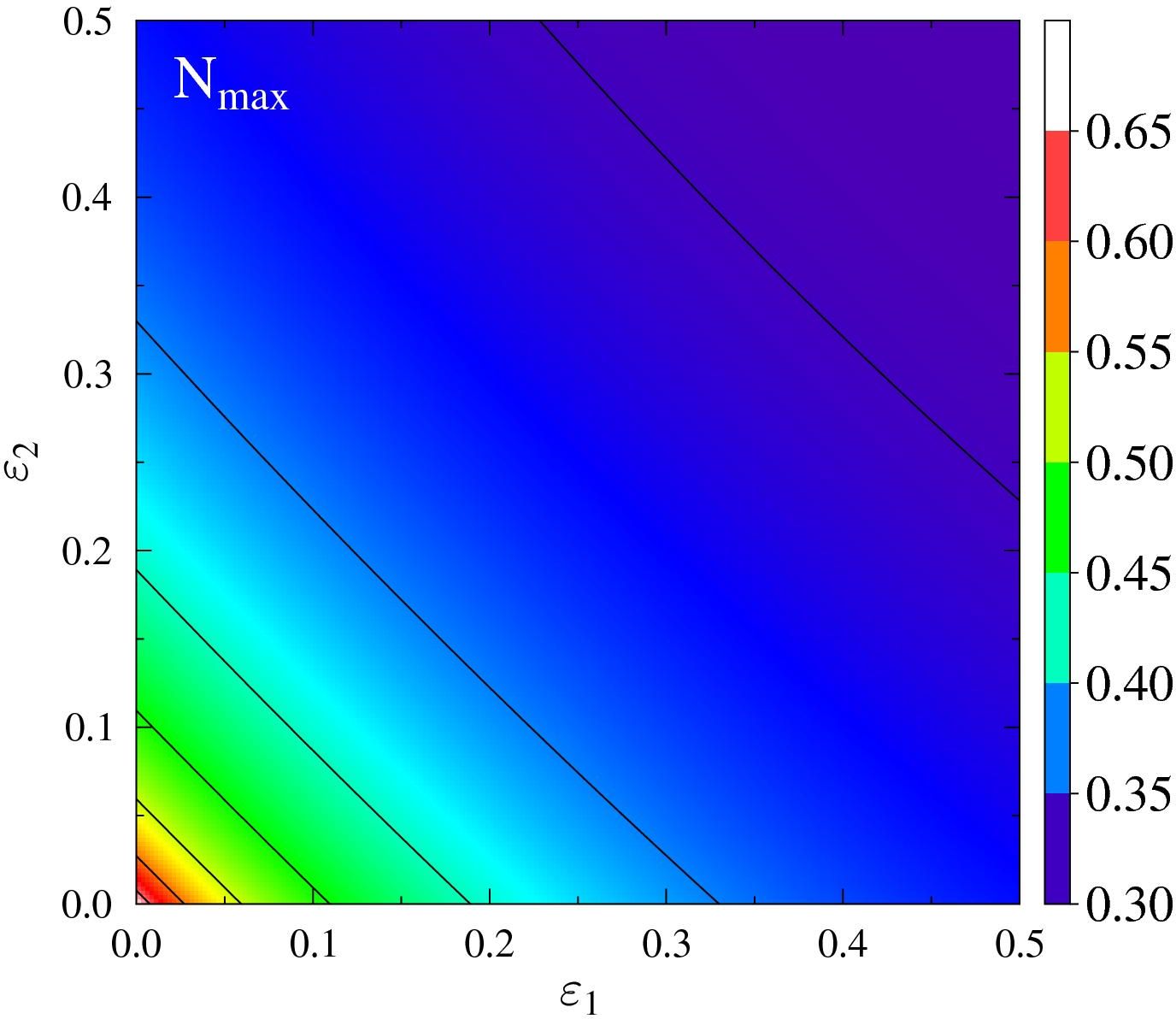}
\caption{Left and middle: The optimal couplings $\lambda_{i{\rm (opt)}}$
	for the first and second quantum dot, respectively,
	as functions of QDs energies $\varepsilon_{1}$ and $\varepsilon_{2}$, corresponding to the maximum negativity.
	Right: The maximal negativity $N_{\rm max}$ as a function of quantum dot
	energies  $\varepsilon_{1}$ and $\varepsilon_{2}$ determined at the optimal couplings $\lambda_{i{\rm (opt)}}$. Parameters are in units of $\varepsilon_M=1$.
}
\label{fig2}
\end{figure*}

\section{Entanglement negativity}
Below, we employ the fermionic version of the density matrix partial transpose. In particular, we transpose the states of the first QD 
($n_{d_1}$, $n'_{d_1}$) and refer to this operation as $T^f_{n_{d_1}}$.
The logarithmic negativity of a partially transposed density matrix, which is defined in analogy to the separability criterion based on the negative eigenvalues
of partial transpose, is generally considered as the most practical method to capture the entanglement in a composite quantum system comprising more than two subsystems, cf. Refs.~\cite{PhysRevA.65.032314,PhysRevB.95.165101} and references therein. In particular, the logarithmic negativity provides an upper bound on the amount of distillable entanglement in multipartite systems~\cite{PhysRevD.109.L071501}.
We calculate the {\it entanglement negativity} following Refs.~\cite{PhysRevB.95.165101,PhysRevD.109.L071501}.
Using the formula
\begin{eqnarray}\label{partial transpose}
\left(\ket{n'_{d_1},n'_{d_2}}\bra{n_{d_1}, n_{d_2}}\right)^{T^f_{n_{d_1}}}=
(-1)^{\alpha}\ket{n_{d_1},n'_{d_2}}\bra{n'_{d_1}, n_{d_2}},\hspace{8mm}
\end{eqnarray}
with the exponent defined as 
\begin{eqnarray}
\nn
\alpha &=& n_{d_2}n'_{d_2} + n'_{d_1}n'_{d_2}+n_{d_1}n_{d_2}+(n_{d_1}+n_{d_2})(n'_{d_1}+n'_{d_2})\\
& + &\frac{n'_{d_1}(n'_{d_1}+2)}{2}+\frac{n_{d_1}(n_{d_1}+2)}{2},
\end{eqnarray}
we obtain the partially transposed reduced density matrix
\begin{eqnarray}\label{Matrixreduced-2}
\hat\rho_R^{T^f_{n_{d_1}}}=\begin{bmatrix}
\rho_{11} & -i\rho_{21} & \rho_{13} & i\rho_{23} \\
-i\rho_{12} & \rho_{22} & i\rho_{14} & \rho_{24} \\
\rho_{31} & i\rho_{41} & \rho_{33} & -i\rho_{43} \\
i\rho_{32} & \rho_{42} & -i\rho_{34} & \rho_{44}
\end{bmatrix}.
\end{eqnarray}
The logarithmic entanglement negativity is given by
\begin{eqnarray}\label{logarithmic negativity}
\mathcal{N}=\ln\left(\text{Tr}\sqrt{\hat\rho_R^{T^f_{n_{d_1}}}\left(\hat\rho_R^{T^f_{n_{d_1}}}\right)^\dag}\right).
\end{eqnarray}
Our aim is to find the parameters $\varepsilon_M,\varepsilon_{1},\varepsilon_2, \lambda_{1},\lambda_2$
for which the negativity is non-zero. When it concerns not a general but a ground state,
the $C_n$ coefficients split into two subsets 
\begin{subequations}
\begin{equation}
C^{-}=(C_1, C_2, C_3, C_4, 0, 0, 0, 0)
\end{equation}
for $\varepsilon_{1\u},\varepsilon_{2\u}<0$, and correspondingly,  
\begin{equation}
C^{+}=(0, 0, 0, 0, C_5, C_6, C_7, C_8)
\end{equation}
\end{subequations}
otherwise. For example, the non-zero matrix elements of the reduced density matrix for $C^+$ read 
\begin{eqnarray}
\nn&\displaystyle
\rho_{11}=|C_5|^2,\; \rho_{22}=|C_7|^2,\; \rho_{33}=|C_6|^2,\rho_{44}=|C_8|^2,&\\ 
\nn&\displaystyle
\; \rho_{14}=\rho^*_{41}=C_5C_8^*,\; \rho_{23}=\rho^*_{32}=C_7C_6^*,&
\end{eqnarray}
those of $C^-$ are found analogously. Then, we deduce:
\begin{eqnarray}\label{logarithmic negativitytwo}
\hat\rho_R^{T^f_{n_{d_1}}}\left(\hat\rho_R^{T^f_{n_{d_1}}}\right)^\dag_\pm=\begin{bmatrix}
\alpha & 0 & 0 & -iv \\
0 & \beta & -i\mu & 0 \\
0 & i\mu  & \gamma_0 & 0 \\
iv & 0 & 0 & \delta
\end{bmatrix},
\end{eqnarray}
where we have introduced the notations: 
\begin{eqnarray}
\nn
\alpha=a^2+e^2,\; \beta=b^2+f^2,\; \gamma_0=c^2+f^2,  \\
\nn
\delta=e^2+d^2,\; \mu=(b-c)f,\; v=(a-d)e.
\end{eqnarray}
For the case $(+)$ the latin symbols are
\begin{eqnarray}
\nn
a=|C_5|^2,\; b=|C_7|^2,\; c=|C_6|^2,\\
\nn
d=|C_8|^2,\; e=C_6C_7,\; f=C_5C_8 , 
\end{eqnarray}
while for the case $(-)$ they are 
\begin{eqnarray}
\nn
a=|C_1|^2,\; b=|C_3|^2,\;  c=|C_2|^2, \\
\nn
d=|C_4|^2,\; e=C_2C_3,\; f=C_1C_4. 
\end{eqnarray}
The eigenvalues of the matrix Eq.~(\ref{logarithmic negativitytwo}) 
\begin{subequations}
\begin{eqnarray}\label{The eigenvalues}
&&r_{1,2}=\frac{1}{2}\left(\beta+\gamma_0\pm\sqrt{(\beta-\gamma_0)^2+4\mu^2}\right),\\
&&r_{3,4}=\frac{1}{2}\left(\alpha+\delta\pm\sqrt{(\alpha-\delta)^2+4\nu^2}\right),
\end{eqnarray}
\end{subequations}
are evaluated for dimensionless units, assuming the overlap between Majorana modes  $\omega=\varepsilon_M/2=1$.
Then, the model's set of independent parameters is reduced to $\varepsilon_{1},\varepsilon_2$ and $\lambda_{1},\lambda_2$. 

\begin{figure*}
\includegraphics[width=0.32\textwidth]{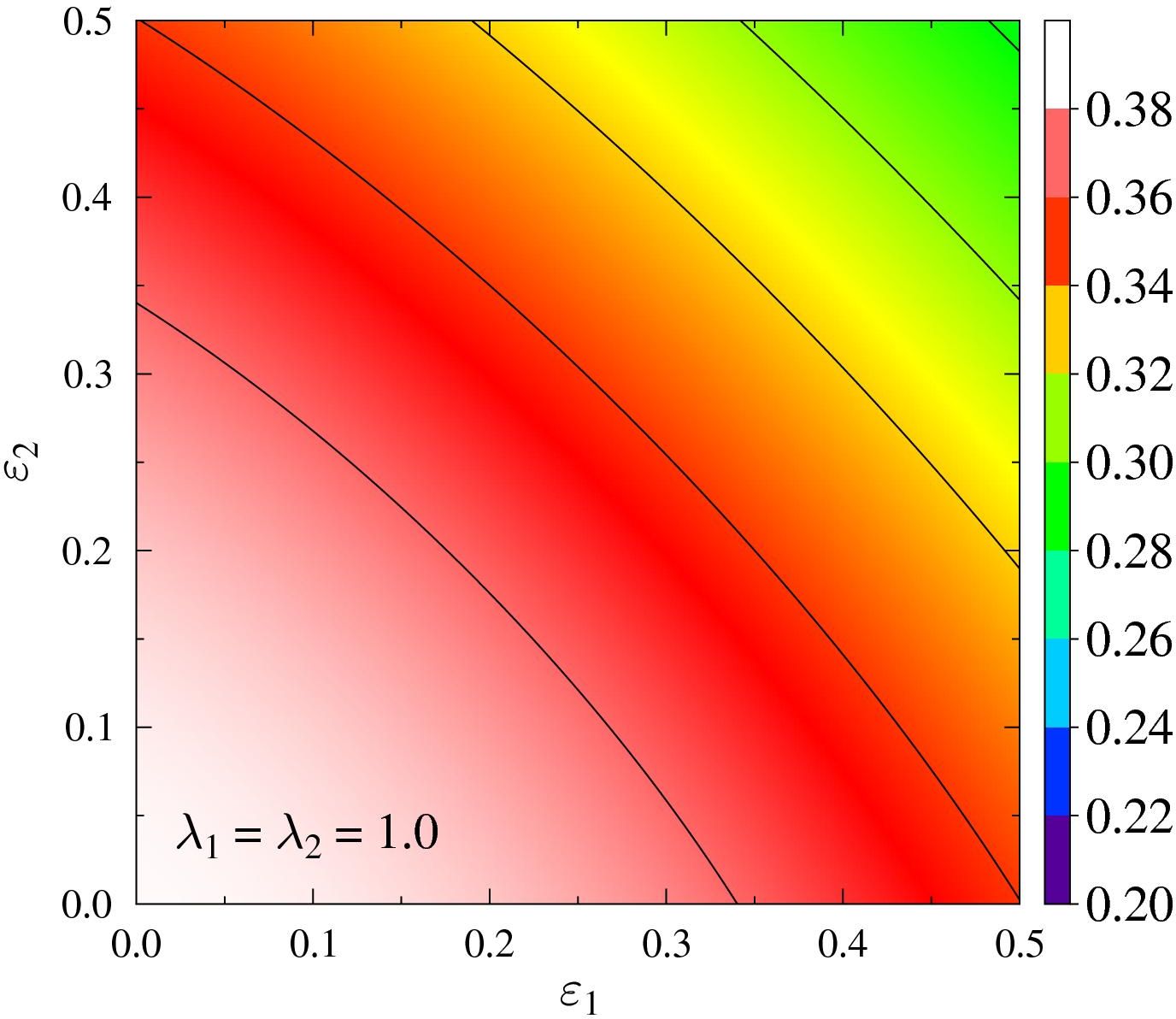}
\includegraphics[width=0.32\textwidth]{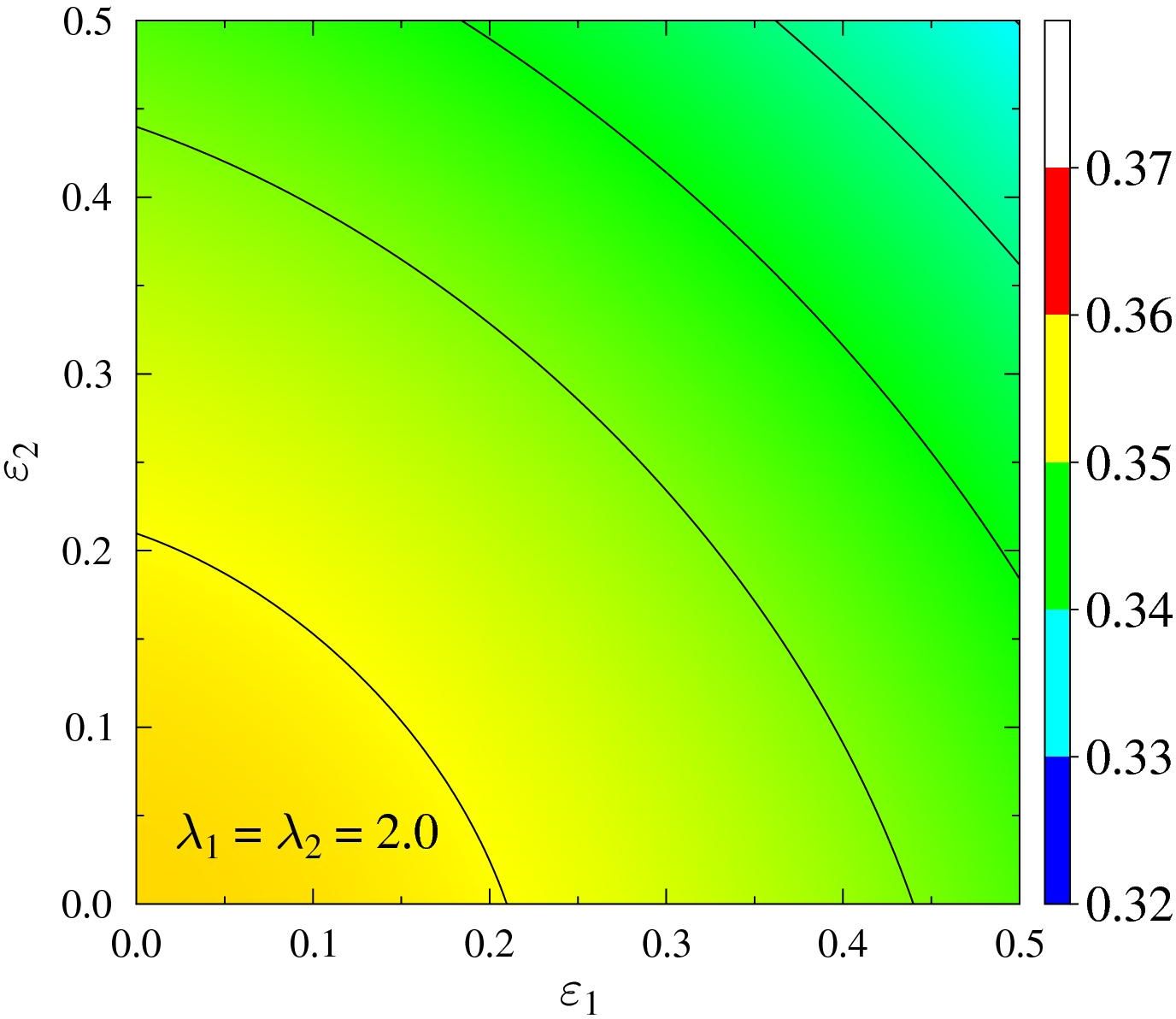}
\includegraphics[width=0.32\textwidth]{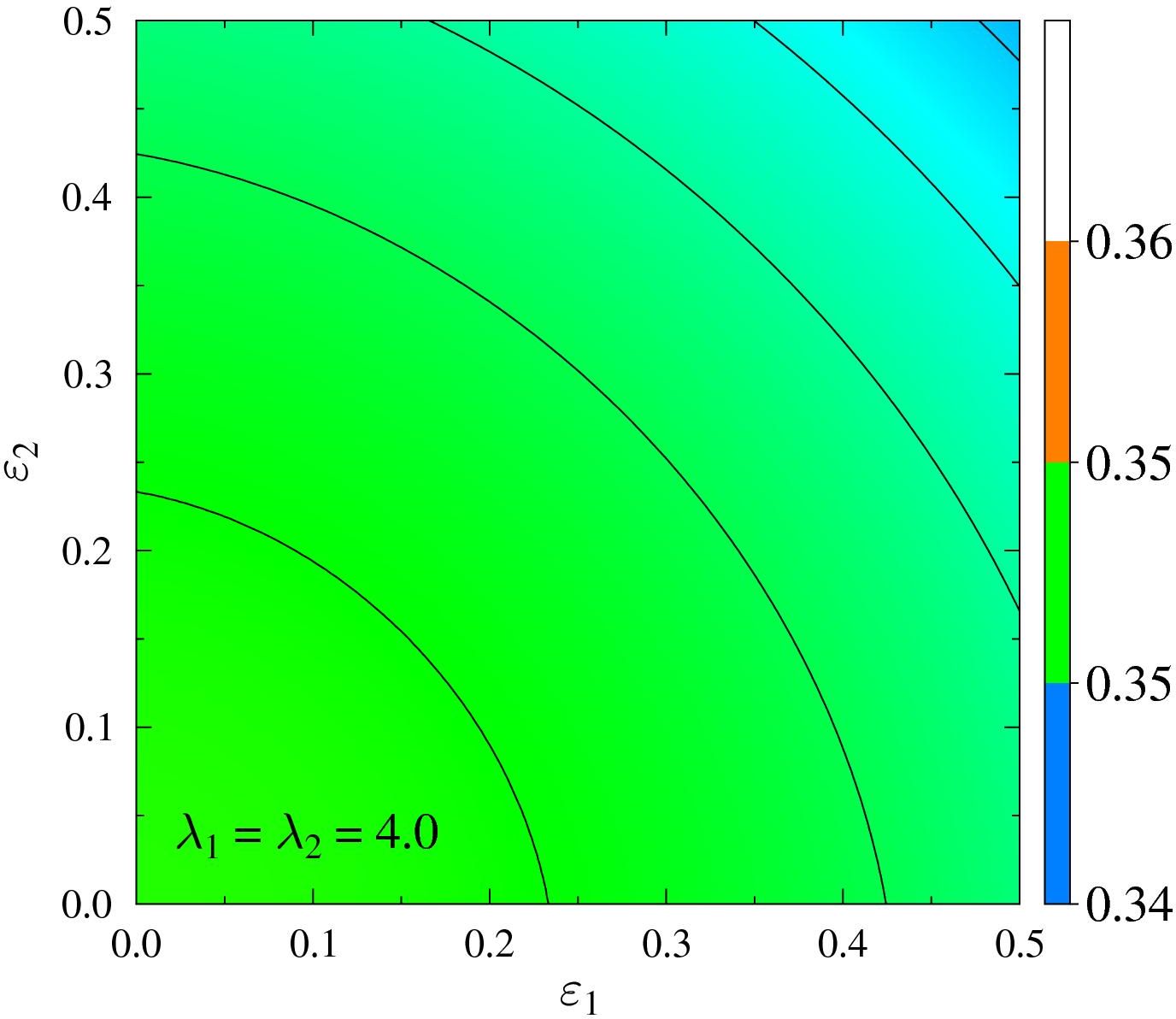}
\caption{
The optimal entanglement determined by the position of the logarithmic negativity maximum
taken at fixed values of $\lambda^{}_1=\lambda^{}_2$ within the plane
spanned by the quantum dots' energies $\{\varepsilon^{}_1,\varepsilon^{}_2\}$.
All quantities are in units of the Majorana overlap $\omega=\varepsilon_M/2$.
}
\label{fig2sup}
\end{figure*}

\section{Optimal entanglement}

In the left (middle) panel of Fig.~\ref{fig2}
we present the optimal value for the coupling $\lambda_{1(\rm opt)}$ [$\lambda_{2(\rm opt})$]
between the first (second) quantum dot and the respective Majorana mode,
corresponding to the maximal negativity, as a function of quantum dot energies $\varepsilon_{1}$ and $\varepsilon_{2}$.
As one can see, the optimal coupling parameters increase with QDs energies,
indicating that a stronger coupling is needed to reach a sensible entanglement.
Hence, the entanglement is mostly pronounced at lower energies,
compared to the overlap energy of the Majorana mode  $\omega=\varepsilon_M/2$.
An interesting observation is that the entanglement degree does not depend
on the QDs energies $\varepsilon_{1}$ and $\varepsilon_{2}$ separately,
but rather on their sum $\varepsilon_{1}+\varepsilon_{2}$.
Therefore, the counter-lines of the same negativity
form the isosceles right triangles, as can be seen in the right panel of Fig.~\ref{fig2}.

In Fig.~\ref{fig2sup}, the optimal entanglement is depicted as maxima of the logarithmic negativity for fixed symmetric QDs-Majorana coupling strength $\lambda^{}_1=\lambda^{}_2$ but changing the quantum dots energies $\varepsilon_1$ and $\varepsilon_2$.
These dependencies represent a family of concentric circle-like curves,
while the curvature is stronger for smaller QDs energies $\varepsilon_{1}$, $\varepsilon_{2}$
and evolves to almost straight lines at larger ones.
Moreover, the curvature strongly depends on
the strength of coupling between the quantum dots and Majorana modes.
As can be seen, the stronger the coupling, the larger curvature is observed,
cf. e.g. the right panel in Fig.~\ref{fig2sup}.

\begin{figure*}
\includegraphics[width=7cm]{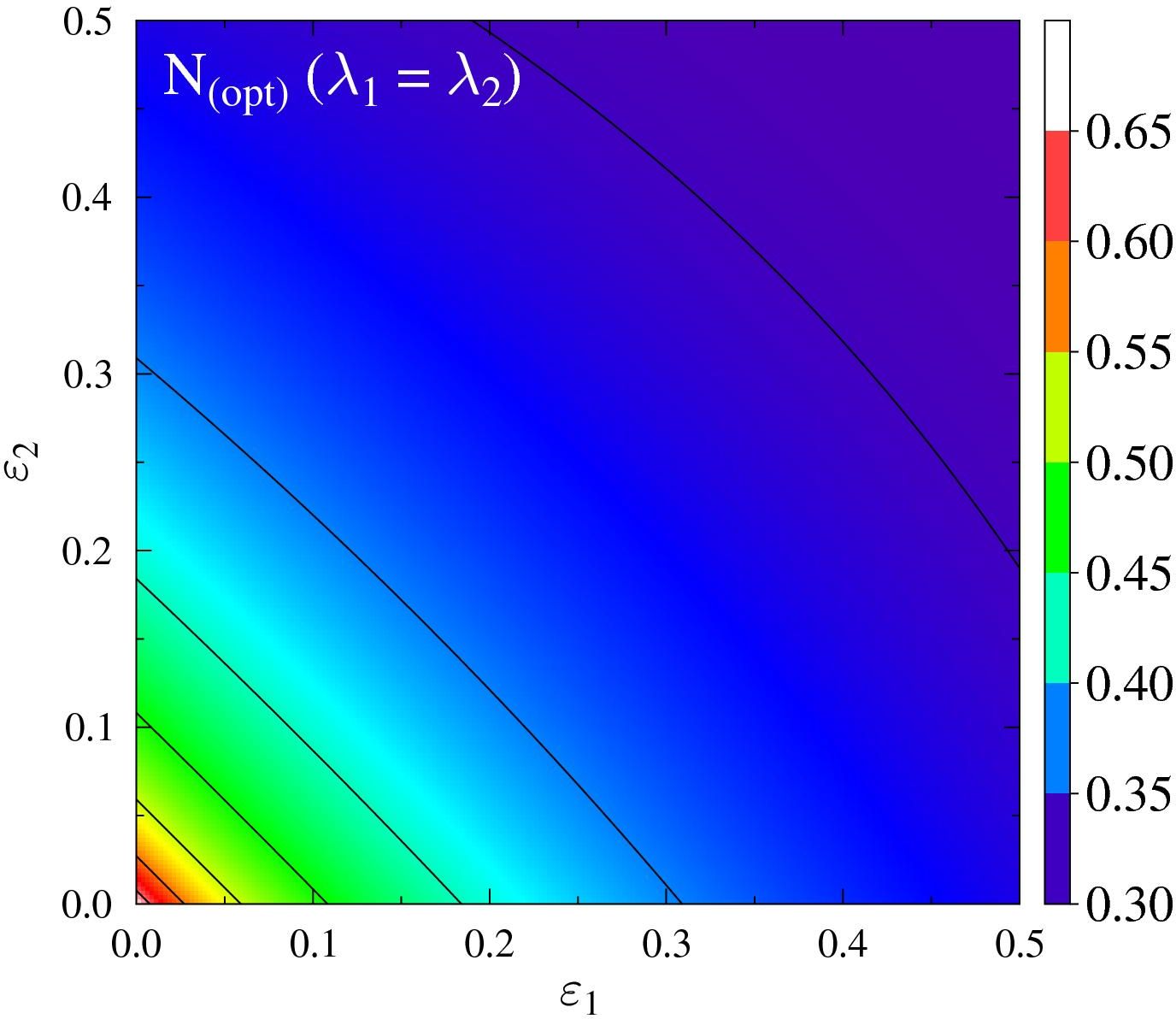}
\hspace{5mm}
\includegraphics[width=6.8cm]{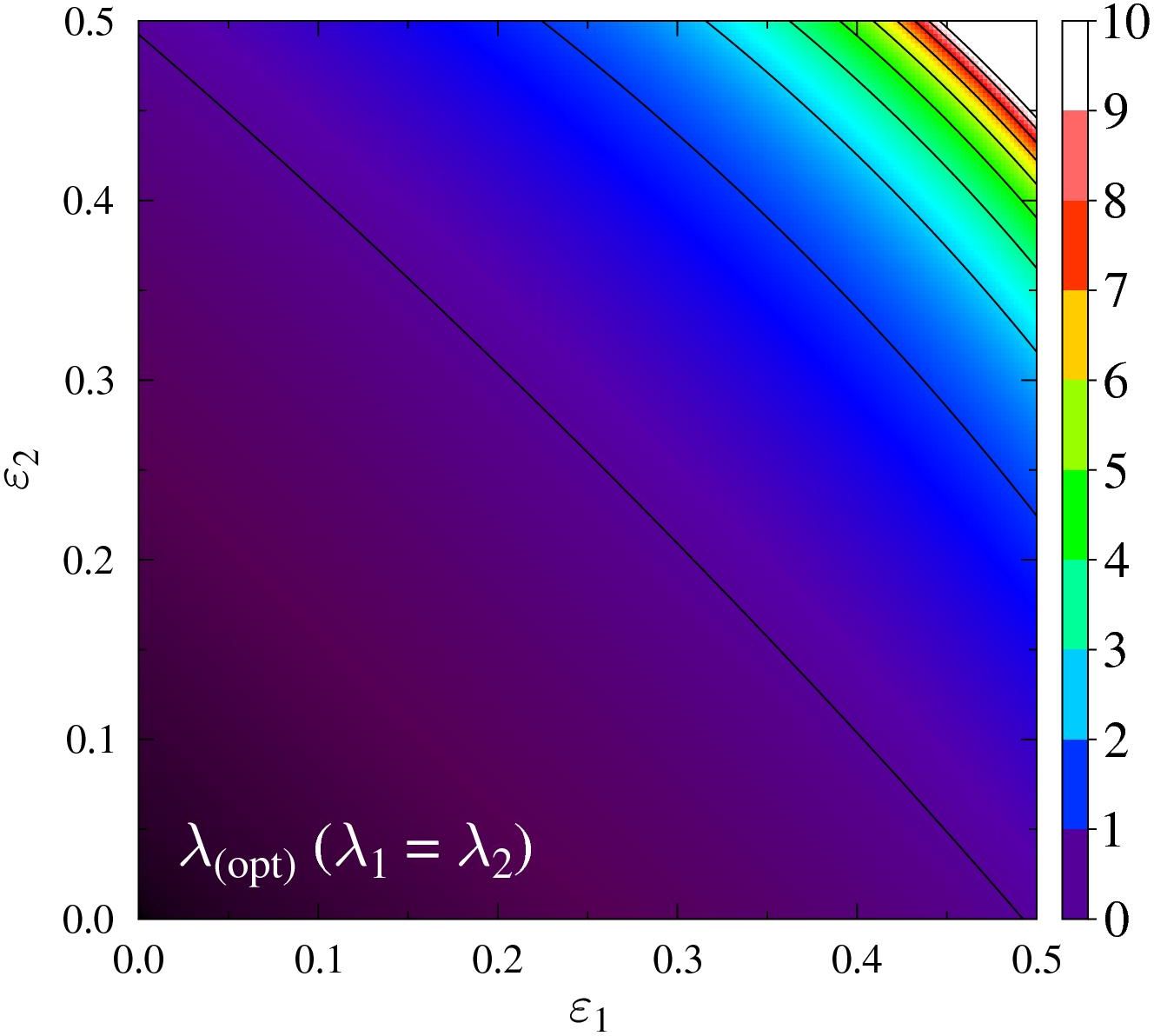}
\caption{
Left: The optimal logarithmic negativity calculated for the equal coupling strengths $\lambda_1=\lambda_2$
as a function of quantum dot energies.
Right: The corresponding optimal coupling $\lambda_{\rm (opt)}$ ($\lambda_1=\lambda_2$)
as function of quantum dot energies $\varepsilon^{}_1$ and $\varepsilon^{}_2$.
}
\label{fig2new}
\end{figure*}

Additionally, in Fig.~\ref{fig2new} we present the optimal logarithmic
negativity for symmetric couplings between quantum dots
and Majorana modes $\lambda_1=\lambda_2$
as function of the QDs energies, together with the corresponding optimal coupling $\lambda_{\rm (opt)}$.
Similarly as before, one can clearly observe that considerable
maximal negativity develops for small detunings 
of the quantum dot energy levels from resonance.
Since the limit of small energies corresponds to the maximal negativity,
let us inspect this parameter space in more detail.

\begin{figure*}
\includegraphics[width=0.85\textwidth]{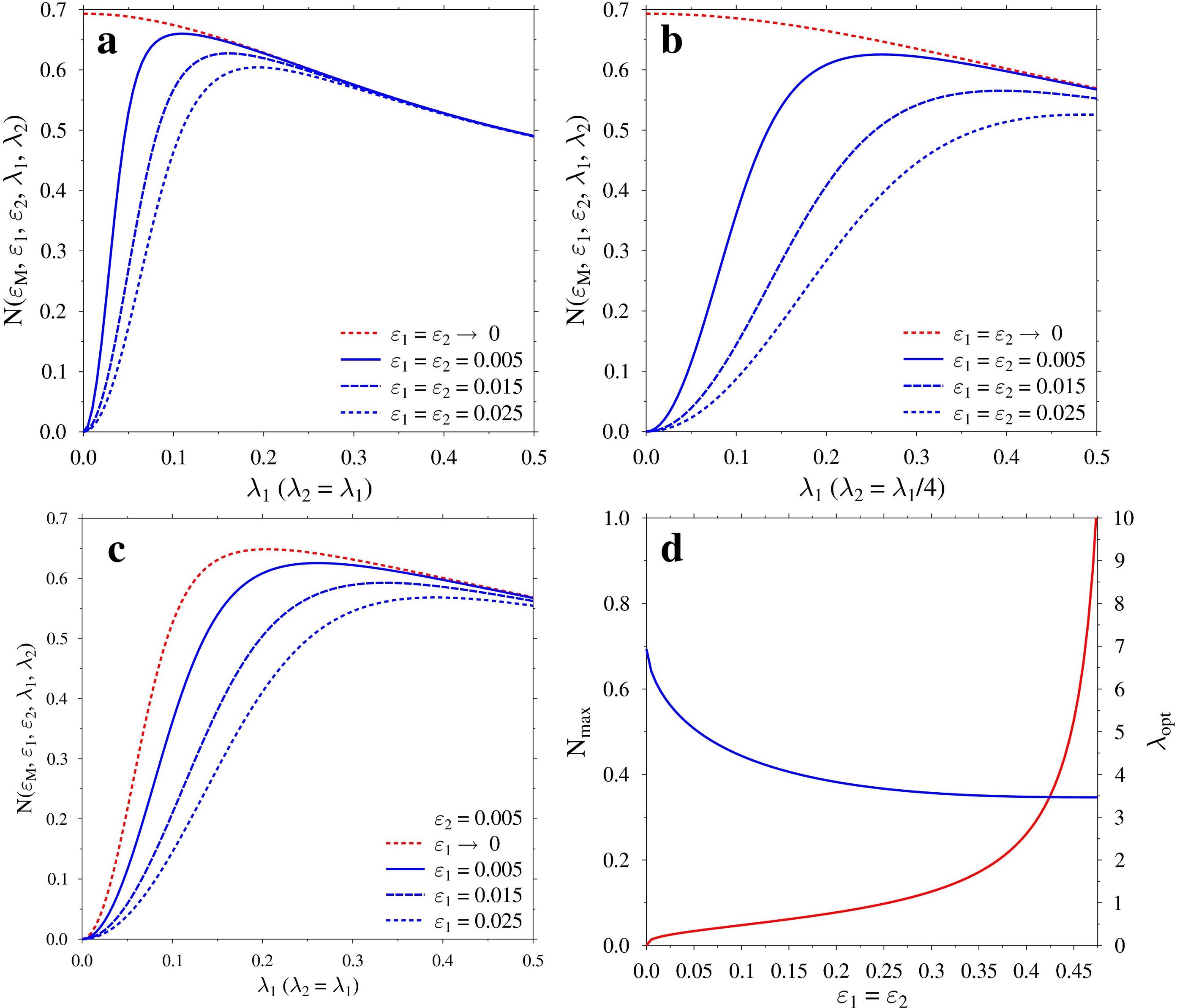}
\caption{(a)-(c) The negativity as a function of the coupling strength $\lambda_{1}=\lambda_{2}$
	in the limit of small QDs energies $\varepsilon_{1}$ and $\varepsilon_2$, as indicated in the legends.
    In (a) and (b) $\varepsilon_{1}=\varepsilon_2$,
    whereas $\varepsilon_2=0.005$ in (c), while $\varepsilon_1$ is tuned.
	The blue line in (d) shows the negativity decay with increasing energies of the dots ($\varepsilon_{1}=\varepsilon_2$), while the red line describes the corresponding optimal value of the coupling strength. All energies are scaled in energy units of the Majorana overlap energy $\omega$.
	}
\label{fig4}
\end{figure*}

In Fig.~\ref{fig4}, we plot the dependence of the negativity in the regime of small QDs energies,
as indicated in the legends, for equal [Figs.~\ref{fig4}(a) and (c)], and unequal [Fig.~\ref{fig4}(b)]
coupling strengths $\lambda_{1}$ and $\lambda_{2}$.
While Figs.~\ref{fig4}(a) and (b) show the case when $\varepsilon_1 = \varepsilon_2$,
Fig.~\ref{fig4}(c) corresponds to the case when $\varepsilon_1$ is changed, while 
$\varepsilon_2$ is fixed.
The red dotted line defines the maximal asymptotic limit that the negativity value never crosses.
It is clearly seen, that the dependence of the negativity on the coupling strength is not monotonic,
but exhibits a particular maximum that moves to larger coupling with increasing energies of the quantum dots.
The blue line in Fig.~\ref{fig4}(d) shows the decay of negativity with increasing DQ energies ($\varepsilon=\varepsilon_1 = \varepsilon_2$),
while the red line describes the corresponding optimal value of the coupling strength. 

\begin{figure*}
\includegraphics[width=5.3cm]{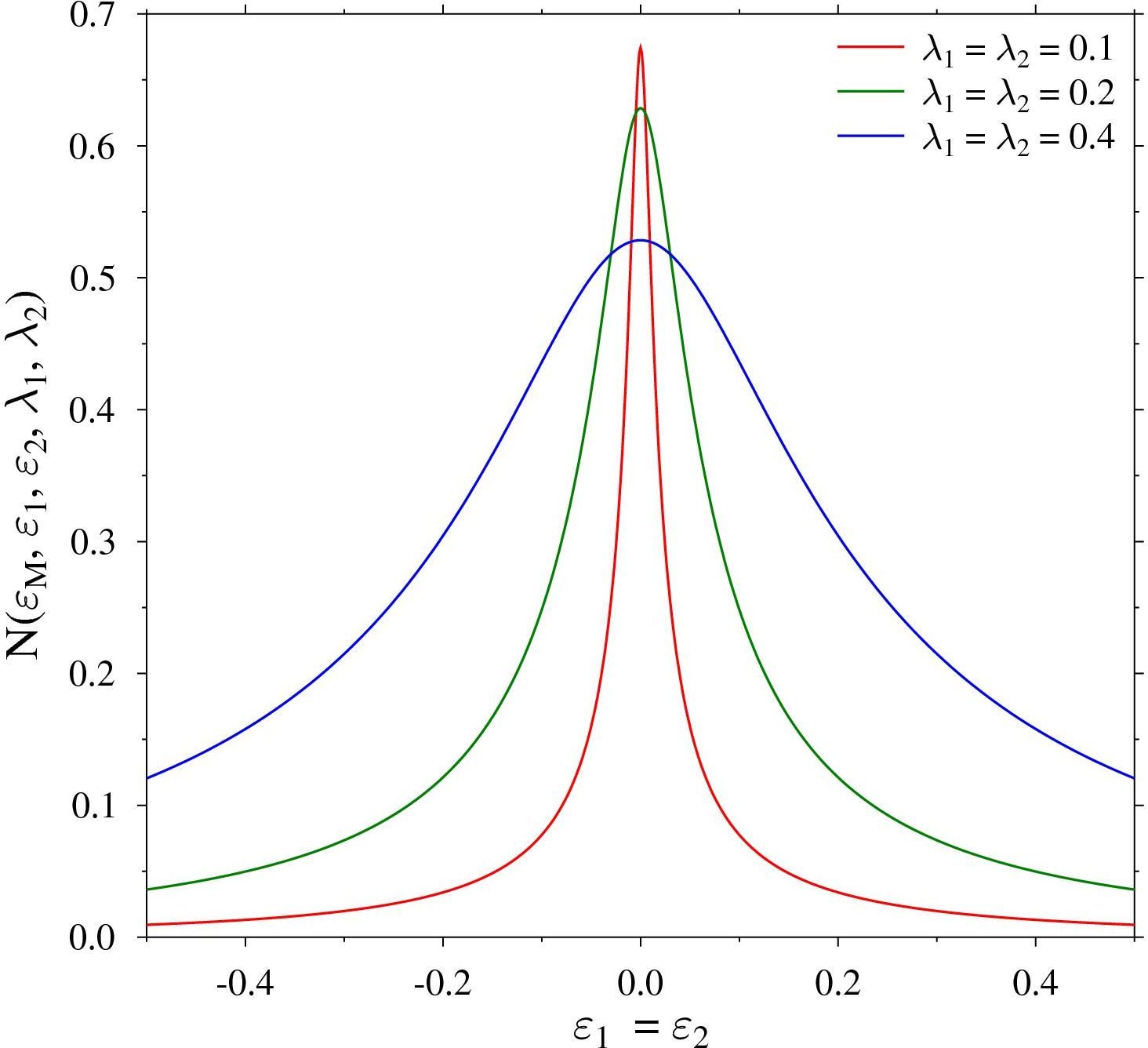}
\includegraphics[width=5.8cm]{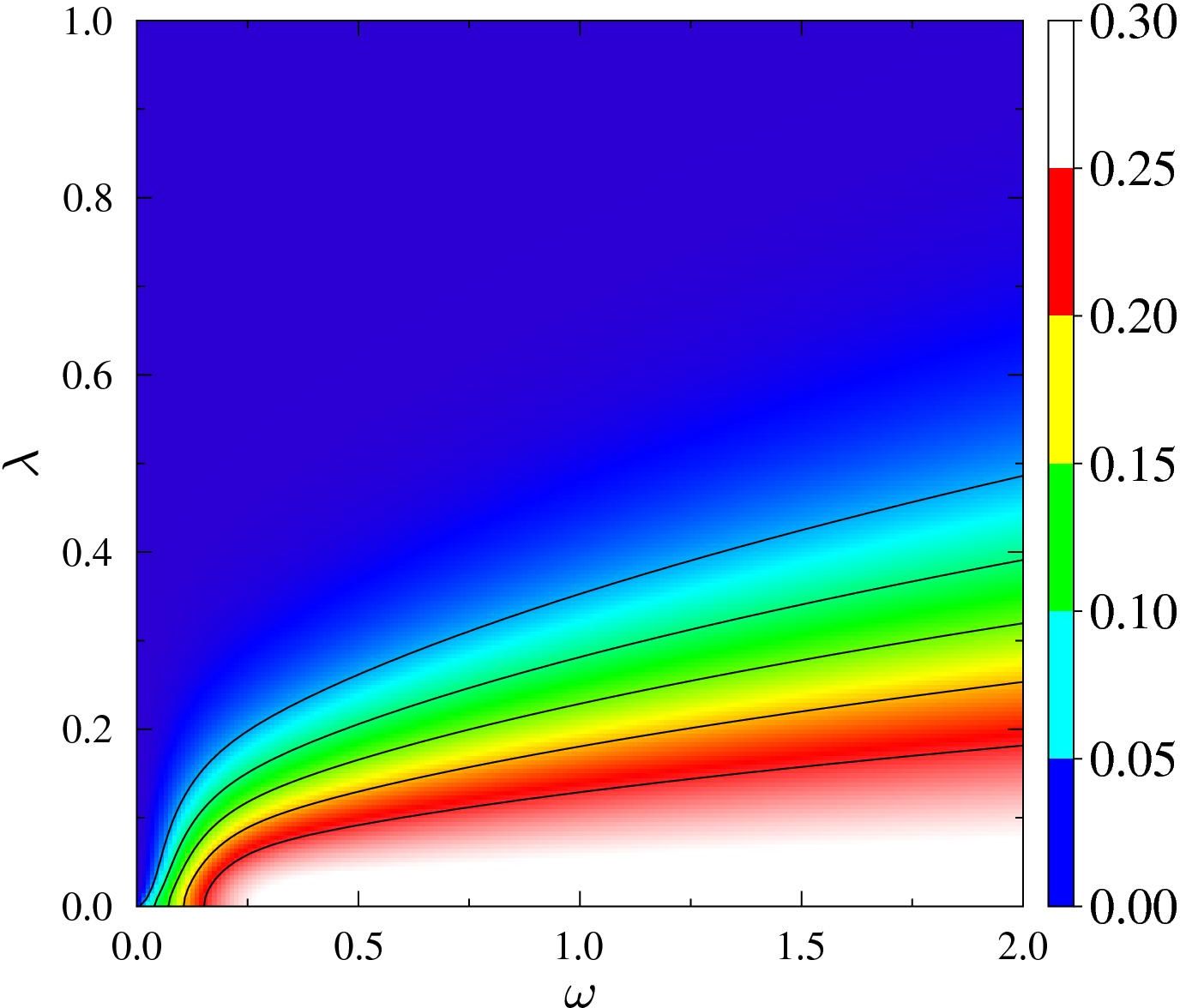}
\includegraphics[width=5.8cm]{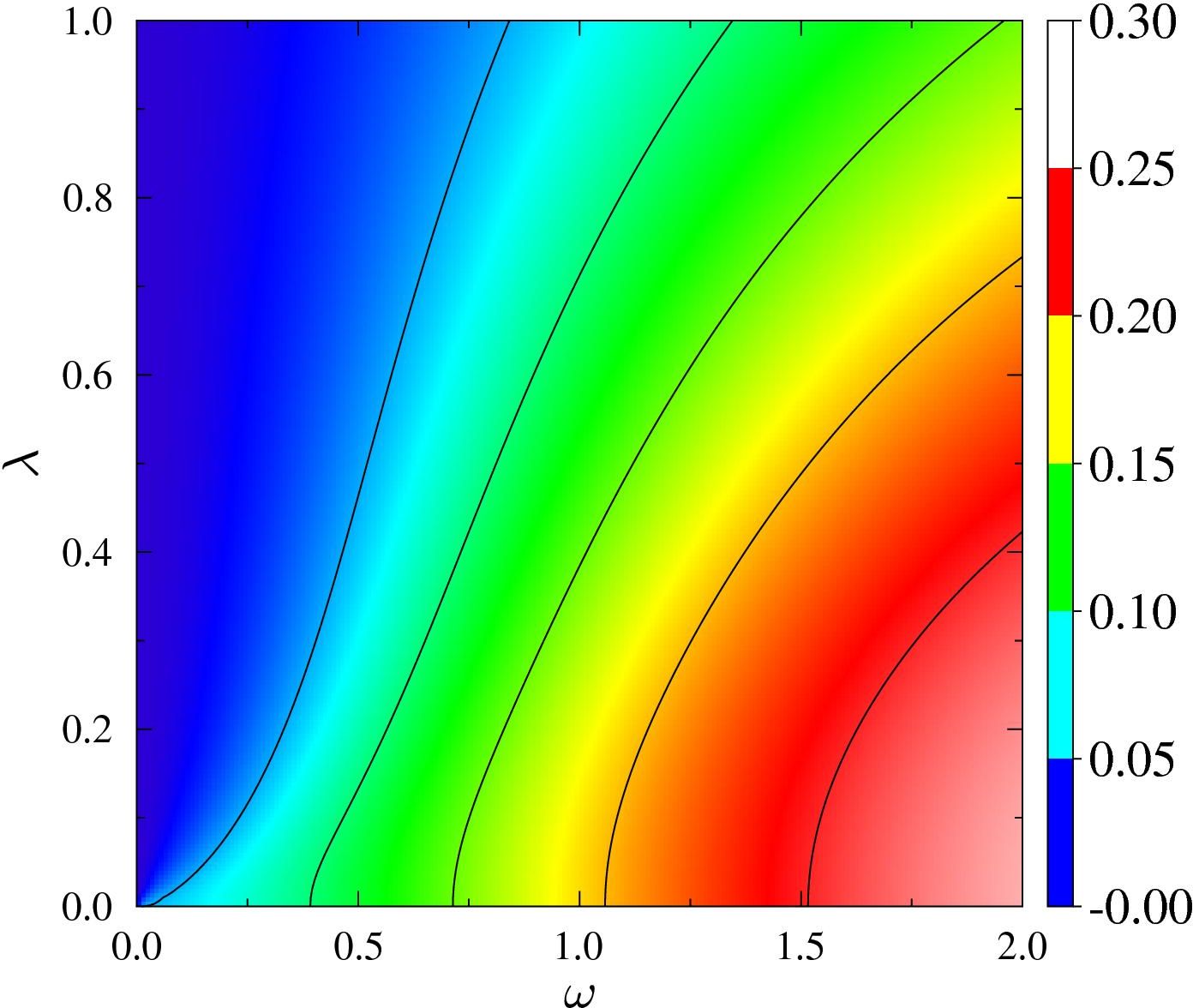}
\caption{
Left: The negativity as a function of the quantum dots energies  $\varepsilon_{1}=\varepsilon_{2}$ (symmetric case) for different values of the 
coupling parameters $\lambda_{1}=\lambda_{2}$ with the Majorana fermions, as indicated.
The negativity is suppressed with increasing the coupling strength $\lambda^{}_{1}=\lambda_2$. 
Middle and right: The quantum mutual information as a function of the coupling strength $\lambda=\lambda_1=\lambda_2$
between the Majorana fermions and electrons and the overlap between the Majorana modes $\varepsilon_M=2\omega$. 
The results are shown for the dimensionless temperature (middle) $T=1$ and (right) $T=10$.
The quantum mutual information between the dots increases with increasing the overlap energy between
the Majorana modes $\omega=\varepsilon_M$ and decreases with the growing ratio of $\lambda/\omega$.
}
\label{fig5}
\end{figure*}

The same negativity, although now plotted as a function of the QD energies $\varepsilon_{1}=\varepsilon_{2}$
at fixed coupling strengths $\lambda_1=\lambda_2$, is shown in the left panel of Fig.~\ref{fig5}.
Here we observe the entanglement suppression, if one of the parameters increases beyond the optimal value. 
Negativity takes the maximum value for $\varepsilon_1 = \varepsilon_2 = 0$,
which decreases with increasing energies of the quantum dots and/or the strength of coupling
to Majorana mode.

\section{Concurrence}
A pure state $\ket{\psi}$ of any bipartite quantum system can be written in a “magic basis”, constructed from the Bell states with extra prefactors $\ket{\psi}=\alpha_i\ket{b_i}$, where $\ket{b_1}=1/\sqrt{2}\ket{\Phi^+}$, $\ket{b_2}=i/\sqrt{2}\ket{\Phi^-}$, $\ket{b_3}=i/\sqrt{2}\ket{\Psi^+}$, $\ket{b_4}=1/\sqrt{2}\ket{\Psi^-}$ in standard notation for the Bell states of a generic bipartite system, cf. Ref.~ \cite{PhysRevLett.78.5022}. Then, the entanglement of a pure state can be quantified via the measure known as the {\it concurrence} $\mathcal{C}(\ket{\psi})=|\sum\alpha_i^2|$. The concurrence can also be generalized for mixed (e.g., thermal) states \cite{PhysRevA.64.042302}. To explore the temperature dependence of the entanglement, we consider the thermal concurrence. Particularly, we choose the parameter set similar to that employed in Ref.~\cite{PhysRevB.110.224510}:  $\varepsilon^{}_M = 2\omega$, 
$\varepsilon^{}_{1}=\varepsilon^{}_{2}=0$, $\lambda^{}_1=-\lambda^{}_2=\sqrt{2}\lambda$. 
Then, the energy eigenbasis of the system reads:   
\begin{subequations}
\begin{eqnarray}
\displaystyle\ket{e_1}&=&-\eta_+\ket{1_f}\otimes\ket{\Phi_d^-}+\zeta_+\ket{0_f}\otimes\ket{\Psi_d^+}, \\
\displaystyle\ket{e_2}&=& \eta_-\ket{1_f}\otimes\ket{\Phi_d^-}+\zeta_-\ket{0_f}\otimes\ket{\Psi_d^+}, 
\end{eqnarray}
for the eigenvalue $E_1=-\sqrt{\omega^2+4\lambda^2}$; 
\begin{eqnarray}
\ket{e_3}&=&-\ket{0_f}\otimes\ket{\Psi_d^-},\\
\ket{e_4}&=&-\ket{0_f}\otimes\ket{\Phi_d^-},
\end{eqnarray}
for the eigenvalue $E_2=-\omega$; 
\begin{eqnarray}
\ket{e_5} &=& \ket{1_f}\otimes\ket{\Phi_d^+}, \\
\ket{e_6} &=& \ket{1_f}\otimes\ket{\Psi_d^+},
\end{eqnarray}
for the eigenvalue $E_3=\omega$; and finally
\begin{eqnarray}
\ket{e_7}&=&-\eta_-\ket{1_f}\otimes\ket{\Phi_d^-}+\zeta_-\ket{0_f}\otimes\ket{\Psi_d^+}, \\
\ket{e_8}&=&\eta_+\ket{0_f}\otimes\ket{\Phi_d^+}+\zeta_+\ket{1_f}\otimes\ket{\Psi_d^-},
\end{eqnarray}
\end{subequations}
for the eigenvalue $E_4=\sqrt{\omega^2+4\lambda^2}$. Here $\ket{\Psi^+_d}$, $\ket{\Phi^-_d}$ are the Bell states of the quantum dots and the shorthands: 
\begin{subequations}
\begin{eqnarray}
\eta_\pm &=&\frac{2\lambda}{\sqrt{4\lambda^2+(\omega\pm\Delta)^2}},\\ 
\zeta_\pm &=&\frac{\omega\pm\Delta}{\sqrt{4\lambda^2+(\omega\pm\Delta)^2}} ,
\end{eqnarray}
\end{subequations}
with $\Delta = \sqrt{\omega^2 + 4\lambda^2}$.

The reduced density matrix of the quantum dot subsystem reads
\begin{eqnarray}\label{thermal one}
\hat\rho_d&=&Z^{-1}\bigg(\Phi_+(T)\ket{\Phi_d^+}\bra{\Phi_d^+}+\Phi_-(T)\ket{\Phi_d^-}\bra{\Phi_d^-}\nonumber\\
&+&\Psi_+(T)\ket{\Psi_d^+}\bra{\Psi_d^+}+\Psi_-(T)\ket{\Psi_d^-}\bra{\Psi_d^-}\bigg),
\end{eqnarray}
where
\begin{equation}
Z=2\sum\limits_{n=1}^4e^{-E_n/T},
\end{equation}
and
\begin{subequations}
\begin{eqnarray}\label{thermal one plus}
\Phi_-(T)&=&(\eta_+^2+\eta_-^2)e^{-E_1/T}+e^{-E_2/T}+\eta_-^2e^{-E_4/T},\hspace{7mm}\\
\Psi_+(T)&=&(\zeta_+^2+\zeta_-^2)e^{-E_1/T}+e^{-E_3/T}+\zeta_-^2e^{-E_4/T},\hspace{7mm}\\
\Phi_+(T)&=&e^{-E_3/T}+\eta_+^2e^{-E_4/T},\\
\Psi_-(T)&=&e^{-E_2/T}+\zeta_+^2e^{-E_4/T}.
\end{eqnarray}
\end{subequations}
Then, following Ref.~\cite{PhysRevA.64.042302}, the thermal concurrence of the state $\hat\rho^{}_d$ should be calculated in the computational basis $\ket{00},\ket{01},\ket{10},  \ket{11}$ through the following formula $C=\text{max}\left\lbrace \lambda_1-\lambda_2-\lambda_3-\lambda_4,0\right\rbrace $,
where $\lambda_n$ are the square roots of the eigenvalues of the matrix 
\begin{equation}
R=\rho_d(\hat\sigma_y\otimes\hat\sigma_y)\rho_d^*(\hat\sigma_y\otimes\hat\sigma_y)
\end{equation}
in decreasing order. Taking into account Eq.~(\ref{thermal one}), we obtain four eigenvalues 
\begin{equation}
\frac{1}{4Z^2}
\left\lbrace\Phi^2_\pm(T), \Psi^2_\pm(T) \right\rbrace. 
\end{equation}
To calculate the \textit{concurrence} we need to order these
eigenvalues in decreasing order. The concurrence is a function of $\{\omega,\lambda, T\}$.
We are particularly interested in low temperature and strong coupling limit $T<\omega,\lambda_{1},\lambda_2$,
when the dominant exponent in  Eq.~(\ref{thermal one plus}) is $e^{-E_1/T}$.
Then, the expression for the density matrix simplifies to
\begin{eqnarray}\label{thermal final}
\hat\rho_d=\frac{1}{2}(\eta_+^2+\eta_-^2)\ket{\Phi_d^-}\bra{\Phi_d^-}+
(\zeta_+^2+\zeta_-^2)
\ket{\Psi_d^+}\bra{\Psi_d^+},
\hspace{5mm}
\end{eqnarray}
and for the concurrence we deduce analytic result
\begin{eqnarray}\label{concurrence we deduce analytic}
\mathcal{C}=\vert 1-\eta_+^2-\eta_-^2\vert.
\end{eqnarray}
From Eq.~(\ref{concurrence we deduce analytic}), it is easy to see that the concurrence is maximal for $\omega>\lambda>T$,  $\mathcal{C}=1-\mathcal{O}(\lambda^2/\omega^2)$ and approaches zero if $\lambda>\omega$. In the high temperature limit, the concurrence is zero as it is confirmed by numerical calculations.

\noindent
One more remark is in order here: For finite temperatures and strong couplings, deviations of $\epsilon^{}_{i=1,2}$ from zero can lead to the hybridization of quantum dots with quasiparticles in the superconductor above the gap in the quantum nanowire. To avoid the measurement pollution due to this hybridization one has to be careful with the variation range of the temperature. The superconducting gap provides a natural effective temperature threshold. 
Measurements at temperatures below this threshold should capture the properties of individual parts of the studied system, 
while at temperatures above the threshold the hybridization effects might become non-negligible and ultimately lead to the destruction of the superconducting state.

\section{Quantum mutual information}
Entropic measures of quantum correlations witness a radical departure from the conditions of the classical physics: While the classical entropy is an additive and extensive quantity, the quantum entropy of a bipartite pure state $\hat\rho_{AB}$ is zero $S(\hat\rho_{AB})=-{\rm Tr}\left(\hat\rho_{AB}\log\hat\rho_{AB}\right)=0$, and the entropy of its subsystem $\hat\rho_A={\rm Tr}_B\left(\hat\rho_{AB}\right)$ is nonzero $S(\hat\rho_{A})=-{\rm Tr}\left(\hat\rho_{A}\log\hat\rho_{A}\right)\neq 0$ if the state $\hat\rho_{AB}$ is entangled.   
The quantum mutual information of a general bipartite system $\hat\rho_{AB}$ is defined as follows \cite{wilde2013quantum,PhysRevA.81.042105}
\begin{eqnarray}\label{QMI general bipartite system}
\mathcal{I}(\hat\rho_{AB})=S(\hat\rho_A)+S(\hat\rho_B)-S(\hat\rho_{AB}).
\end{eqnarray}
In order to get further insights into the entanglement properties,
we calculate the quantum mutual information for the reduced density matrix of QDs $\hat\rho_R$:
\begin{eqnarray}\label{quantum mutual}
\mathcal{I}(\hat\rho_R)=S(\hat\rho_R^{d_1})+S(\hat\rho_R^{d_2})+\sum\limits_{n=1}^4\rho^{}_{R,nn}\log(\rho^{}_{R,nn}),
\end{eqnarray}
where the eigenvalues of the matrix $\rho_{R,nn}$ are given by:
\begin{subequations}
\begin{eqnarray}\label{eigenvaluesR}
&&\rho_{R,11}=\frac{1}{2Z}\left(\Phi_++\Phi_-+|\Phi_+-\Phi_-|\right), \\
&&\rho_{R,22}=\frac{1}{2Z}\left(\Psi_++\Psi_-+|\Psi_+-\Psi_-|\right), \\
&&\rho_{R,33}=\frac{1}{2Z}\left(\Phi_++\Phi_--|\Phi_+-\Phi_-|\right), \\
&&\rho_{R,44}=\frac{1}{2Z}\left(\Psi_++\Psi_--|\Psi_+-\Psi_-|\right),
\end{eqnarray}
\end{subequations}
and entropies of the marginal states $\hat\rho_R^{d_1}={\rm Tr}_{d_2}(\hat\rho_R)$, $\hat\rho_R^{d_2}={\rm Tr}_{d_1}(\hat\rho_R)$ in our case read:  
\begin{subequations}
\begin{eqnarray}\label{marginal states}
 S(\hat\rho_R^{d_1})&=&-(\rho_{R,11}+\rho_{R,22})\log_2(\rho_{R,11}+\rho_{R,22})\;\;\nonumber\\
&&-(\rho_{R,33}+\rho_{R,44})\log_2(\rho_{R,33}+\rho_{R,44}),\\
 S(\hat\rho_R^{d_2})&=&-(\rho_{R,11}+\rho_{R,33})\log_2(\rho_{R,11}+\rho_{R,33})\;\;\nonumber\\
&&-(\rho_{R,22}+\rho_{R,44})\log_2(\rho_{R,22}+\rho_{R,44}).
\end{eqnarray}
\end{subequations}
In Fig.~\ref{fig5} we plot the quantum mutual information as function of the coupling strength $\lambda$
between Majorana fermions and electrons and overlap between Majorana modes $\varepsilon_M=2\omega$
for the dimensionless temperatures $T=1$ and $T=10$, respectively.
As can be seen, the quantum mutual information between QDs increases
monotonously with growing the overlap between Majorana modes $\omega$,
but simultaneously decreases with the growing ratio of $\lambda/\omega$.

\section{Summary and conclusions}

In this article we have performed detailed studies of nonlocal entanglement between the two quantum dots interconnected through the topological nanowire, hosting Majorana zero-energy modes. In particular, we have considered and evaluated three key quantities, which are customarily used for the entanglement quantification: the logarithmic entanglement negativity, the quantum concurrences, and the mutual information between QDs.  
Using the logarithmic negativity criterion, we have found that the optimal entanglement, i.e. the maximal negativity value under given circumstances, 
occurs when quantum dots energy levels are around
the overlap energies of Majorana modes leaking onto these quantum dots.
Moreover, we considered thermal concurrence and quantum mutual information to analyze the entanglement properties at finite temperatures. Both quantities exhibit similar behaviors. The quantum mutual information and concurrence between QDs increase as function of the overlap between Majorana modes $\omega$.
On the other hand, they decrease with the ratio $\lambda/\omega$. The reason for such behavior is obvious:
When the interaction strength between Majorana modes and individual QDs exceeds by far the Majorana overlap energy, i.e. $\lambda/\omega\ll1$, the Majorana modes cannot entangle two independent QDs, the QDs disentangle and we effectively have a product of two separable states of each quantum dot $\ket{\Omega^{}_{d_1,\gamma_1}}\otimes\ket{\Omega^{}_{d_2,\gamma_2}}$.

From the experimental point of view, the effects described in this article might well lie within the reach of the currently available techniques, routinely employed by the experimentalists working in the field. In principle, the density matrices of tri- and bipartite states of the QD-nanowire-QD system can be reconstructed by the quantum state tomography method developed by Steffen {\it et al.} in Ref.~\cite{Steffen2006}. From the reconstructed states obtained in this way, one can at least in principle extract the entanglement measures. 
Recent experimental progress in the realization of the minimal Kitaev chain in systems of two QDs and a superconducting nanowire reported in Ref.\cite{Dvir2023} 
offers a perfect application platform for that technique.

\section*{Acknowledgements} 
A.S. acknowledges funding by the grant JSF-22-10-0001 of the Julian Schwinger Foundation for Physics Research.
I.W. and T.D. acknowledge support by the National Science Centre (Poland) through the grant No.\ 2022/04/Y/ST3/00061.


\appendix
\section{Discussion of the correlation effects}
\label{app:Coulomb}

Our analytical methods for evaluating  the mutual entanglement between the dots become rather hardly feasible in the presence of the on-site Coulomb interaction $U\hat{n}_{i\uparrow}\hat{n}_{i\downarrow}$, where $\hat{n}_{i\sigma}=\hat{d}_{i\sigma}^{\dagger}\hat{d}_{i\sigma}$ is a number operator of spin-$\sigma$ electrons on $i$-th quantum dot. To capture the correlation effects, we briefly discuss general expectations, which can be inferred from studying the quasiparticle spectra of hybrid structures analogous to our setup. 

Most of the studies of Coulomb repulsion have so far addressed the single Anderson-type impurity attached to the topological superconductor wire.
In particular, the lowest-order mean field approximation predicted the emergence of bowtie or diamond shapes, depending on the ratio between the overlap $\varepsilon_{M}$ and the hybridization strength $\lambda_i$ \cite{Deng-2018}. The more accurate Hubbard-I approximation indicated that optimal conditions for the leakage of Majorana mode onto the correlated quantum dot takes place when tuning the energy $\varepsilon_{i\sigma}$ or $\varepsilon_{i\sigma}+U$ by the gate potential to the Fermi level \cite{Ricco-2019}. Since in practical realizations of superconducting hybrid structures the Coulomb potential is by far larger than the pairing gap, we can expect the Majorana mode to leak onto the $i$-th quantum dot of our setup when either the energy levels $\varepsilon_{i\sigma}$ (with the spectral weight $1-\left< n_{i,-\sigma}\right>$) or the Coulomb satellites $\varepsilon_{i\sigma}+U$ (with probability $\left< n_{i,-\sigma}\right>$) coincide with the Fermi energy.
Additional effects might arise from the influence of external lead electrons through exchange interactions, which can induce the Kondo effect at low temperatures \cite{Lopez-2013,Lutchyn-2014,Weymann-2017,Vernek-2020}. This issue, however, is outside the scope of our present study.

As concerns the hybrid structures consisting of two quantum dots interconnected through the short-length topological superconductor the nonlocal cross-correlations can arise due to formation of the molecular structure (i.e. mutual quasiparticle states appearing with specific spectral weights in the given spin sector of each QD) \cite{Gorski-2024}. Their manifestation shall be observable under non-equilibrium conditions through the crossed Andreev reflections \cite{paper_2dots,paper_3dots,Li-2020} and/or time-resolved measurements \cite{Diniz-2023,Taranko-2024}. Such methods might verify inter-dot feedback effects transmitted through the short topological nanowire, but experimental evidence is missing, and therefore further understanding of the nonlocal cross-correlation phenomena is desirable. Our present evaluation of quantum entanglement between the quantum dots can be a useful starting point towards such studies.


%

\end{document}